\documentclass[useAMS,usenatbib]{mn2e}

\usepackage{epsfig,lscape} 
\usepackage{hyperref} 
\usepackage{natbib}
\usepackage[caption=false]{subfig}
\usepackage{lscape}
\usepackage{textcomp}
\usepackage{gensymb}
\usepackage{xcolor}
\usepackage{amsmath}
\usepackage{comment}
\usepackage{arydshln}
\usepackage{graphics}
\usepackage[caption=false]{subfig}

\newcommand{\mnras}{MNRAS}
\newcommand{\apj}{ApJ}
\newcommand{\araa}{ARA\&A}
\newcommand{\aap}{A\&A}
\newcommand{\aj}{AJ}
\newcommand{\apjl}{ApJL}
\newcommand{\apjs}{ApJS}
\newcommand{\pasp}{PASP}
\newcommand{\ao}{ApOpt}

\title[Chemical properties of the MW's bar]{The chemical properties of the Milky Way's on-bar and off-bar regions: evidence for inhomogeneous star formation history in the bulge} 

\author[J. Lian et al.]{{Jianhui~Lian}$^{1}$\thanks{jianhui.lian@astro.utah.edu}, {Gail Zasowski}$^{1}$, {Sten Hasselquist}$^{1}$, Justus Neumann$^{2}$, \newauthor {Steven R. Majewski}$^{3}$, Roger E. Cohen$^{4}$, 
Jos\'e G. Fern\'andez-Trincado$^{5}$, 
\newauthor Richard R. Lane$^{6}$, 
Pen\'elope Longa-Pe\~na$^{7}$, Alexandre Roman-Lopes$^{8}$\\
\small $^{1}${Department of Physics \& Astronomy, University of Utah, Salt Lake City, UT 84112, USA}\\
\small $^{2}${Institute of Cosmology and Gravitation, University of Portsmouth, Burnaby Road, Portsmouth, UK, PO1 3FX}\\
\small $^{3}${Department of Astronomy, University of Virginia, Charlottesville, VA 22904-4325, USA}\\ 
\small $^{4}${Departamento de Astronomía, Universidad de Concepci\'on, Casilla 160-C, Concepci\'on, Chile}\\
\small $^{5}${Instituto de Astronom\'ia y Ciencias Planetarias, Universidad de Atacama, Copayapu 485, Copiap\'o, Chile}\\
\small $^{6}${Instituto de Astronom\'ia y Ciencias Planetarias de Atacama, Universidad de Atacama, Copayapu 485, Copiap\'o, Chile}\\
\small $^{7}${Centro de Astronomía, Universidad de Antofagasta, Avenida Angamos 601, Antofagasta 1270300, Chile}\\
\small $^{8}${Departamento de F\'isica, Facultad de Ciencias, Universidad de La Serena, Cisternas 1200, La Serena, Chile}\\
}

\begin{document}
\maketitle 

\begin{abstract}
Numerous studies of integrated starlight, stellar counts, and kinematics have confirmed that the Milky Way is a barred galaxy. However, far fewer studies have investigated the bar's stellar population properties, which carry valuable independent information regarding the bar's formation history. Here we conduct a detailed analysis of chemical abundance distributions ([Fe/H] and [Mg/Fe]) in the on-bar and off-bar regions to study the azimuthal variation of star formation history (SFH) in the inner Galaxy.  
We find that the on-bar and off-bar stars at Galactocentric radii $3<r_{\rm GC}<5\;$kpc have remarkably consistent [Fe/H] and [Mg/Fe] distribution functions and [Mg/Fe]--[Fe/H] relation, suggesting a common SFH shared by the long bar and the disc. 
In contrast, the bar and disc at smaller radii ($2 < r_{\rm GC} < 3\;$kpc) show noticeable differences, with relatively more very metal-rich ($\rm [Fe/H] \sim 0.4$) stars but fewer solar abundance stars in the bar. Given the three-phase star formation history proposed for the inner Galaxy in \citet{lian2020c}, these differences could be explained by the off-bar disc having experienced either a faster early quenching process or recent metal-poor gas accretion. 
Vertical variations of the abundance distributions at small $r_{\rm GC}$ suggest a wider vertical distribution of low-$\alpha$ stars in the bar, which may serve as chemical evidence for vertical heating through the bar buckling process. The lack of such vertical variations outside the bulge may then suggest a lack of vertical heating in the long bar. 
\end{abstract}

\begin{keywords}
		The Galaxy: abundances -- The Galaxy: bulge -- The Galaxy: formation -- The Galaxy: evolution -- The Galaxy: stellar content -- The Galaxy: structure.
\end{keywords}

\section{introduction}
Galactic bars are common, elongated stellar structures found at the center of disc galaxies (e.g., \citealt{jogee2004,aguerri2009,buta2015}). 
Although they contribute a minor fraction of the total stellar mass of a galaxy, bars play an important role in 
galaxy evolution and bulge formation \citep{kormendy2004}. 
The Milky Way is also a barred galaxy, with bar structure identified in both light density profiles \citep{blitz1991} and kinematics \citep{binney1991}. 
The inner Milky Way is composed of an elongated bar (buckled inner bar+planar outer bar) inset in the disc, inner halo stars, and possibly a weak spheroidal classical bulge in the center \citep{barbuy2018}. 

{The chemistry of a stellar population's birth gas cloud is imprinted on the population's chemical composition, which thus reflects the formation and enrichment history of earlier stellar generations.}
To infer a galaxy's history from the chemical compositions of its stars, extensive spectroscopic observations of individual stars formed at different epochs are needed. 
Due to high extinction towards the Milky Way's center, inner Galaxy observations were long limited to small samples located at off-plane regions with relatively lower extinction. With these limitations, most previous studies on the chemical properties of the inner Galaxy focused on the spatially-averaged chemical properties of the inner Galaxy. 
 
Early photometric and spectroscopic studies suggested that bulge stars in the Milky Way are generally old and $\alpha$-rich \citep[e.g.,][]{zoccali2003,cunha2006}. 
More recent large spectroscopic surveys that target the bulge, such as ARGOS \citep{freeman2013} and APOGEE \citep{majewski2017}, are now rapidly overcoming the obstacles faced in earlier bulge studies and providing high-quality spectra for unprecedentedly large samples of stars in the inner Galaxy. 
These data reveal a complex {mixture of bulge stellar populations with a wide range of metallicity and complex structure in their elemental abundances} \citep[e.g.,][]{ness2013,garcia2018,rojas2019,zasowski2019,queiroz2020,rojas2020}, kinematics \citep[e.g.,][]{kunder2012,ness2016a,zasowski2016}, and ages \citep[e.g.,][] {bensby2013,bensby2017,schultheis2017,hasselquist2020}.  

Many studies have argued that the bulge is composed of (at least) two primary populations with distinctive $\alpha$ element abundance \citep{babusiaux2010,hill2011,schultheis2017,rojas2019,lian2020c}.  
A star formation quenching \citep{haywood2018,lian2020b} or interruption \citep{chiappini1997,matteucci2019} that bridges the formation of the two main populations is proposed to explain this $\alpha$-dichotomy.   

The extensive coverage of large surveys enables not only studies of the average bulge properties, or even of its gradients, but also an assessment of the connection between the inner Galactic disc and the bar embedded inside of it \citep[e.g.,][]{bovy2019,wegg2019,queiroz2020b}.
Based on APOGEE data, \citet{bovy2019} studied the 
chemistry, age, and kinematics of the bar as well as off-bar disc inside the bar radius on the plane ($r_{\rm GC}<5$kpc and $|z|<0.3\;$kpc). The authors found that the bar tends to contain more old, metal-poor stars than the disc, which was interpreted as evidence of early bar formation in the Milky Way. 
In contrast, \citet{wegg2019} found that the bar at $r_{\rm GC} \sim 3-4$~kpc is more metal-rich than the disc (using spatial and kinematical definitions to classify bar and disc stars; see also \citealt{queiroz2020b}).

To explore the azimuthal variation of stellar chemical compositions in the inner Galaxy, in this paper we present a detailed comparison between the bar and the off-bar disc at different radial and vertical positions. 
The goal of this study is to understand whether and how the formation histories of the Galactic bar and off-bar disc are different. 

\section{Observations}
\subsection{Sample selection}
\label{sec:sample}

\begin{figure}
	\centering
	\includegraphics[width=10cm,viewport=40 0 580 380,clip]{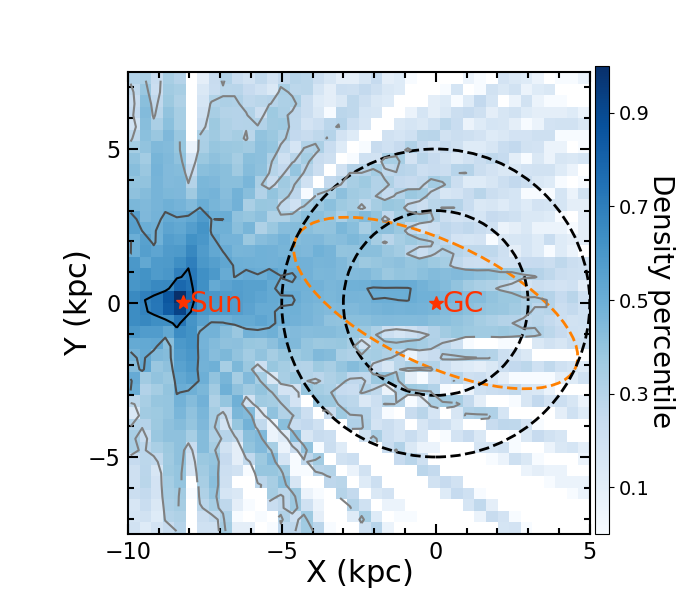}
	\caption{Spatial distribution of APOGEE stars in the Galactic $X-Y$ plane. The larger black circle indicates the total region for our inner Galaxy sample selection, while the smaller circle marks the boundary between radially separated sub-samples (Section~\ref{sec:sample}). The orange ellipse illustrates the bar definition adopted in this work. The positions of the Sun and Galactic center are marked.}
	\label{xy}
\end{figure}

We select stars with APOGEE data in the SDSS-IV Data Release 16 \citep[DR16;][]{ahumada2020,jonsson2020} and post-DR16 APOGEE internal data release, which includes data from observations through March 2020 that have been reduced with a very slightly updated version of the DR16 pipeline (r13).  
APOGEE is a near-infrared, high-resolution spectroscopic survey \citep{blanton2017,majewski2017} that uses custom spectrographs \citep{wilson2019} at the 2.5~m Sloan Telescope and the NMSU 1m Telescope at the Apache Point Observatory \citep{gunn2006,holtzman2010}, and at the 2.5~m Ir\'en\'ee du~Pont telescope \citep{bowen1973} at Las Campanas Observatory.   
APOGEE targets primarily red clump and red giant branch stars throughout the Galaxy
\citep{zasowski2013, zasowski2017}.  

We use chemical abundances and stellar parameters derived by custom pipelines described in \citet{nidever2015} and \citet{garcia2016}\footnote{{An interface for inspecting and downloading APOGEE spectra can be found at \url{https://dr16.sdss.org/infrared/spectrum/search}}} and spectro-photometric distances based on the procedure described in \citet{rojas2017}. We confirm that our results do not change significantly when using StarHorse distances \citep{queiroz2018,queiroz2020}. 
Based on comparison to optical observations, magnesium is shown to be the most reliably measured $\alpha$ element in APOGEE {(see more discussion on the method and reliability of individual elemental abundances in APOGEE in \citealt{jonsson2018,jonsson2020})}. It is therefore used to trace the $\alpha$ abundance in this work. {We note that the elemental [Fe/H] abundance of APOGEE stars is not populated in the catalog when it differs from [M/H] (the total metal content determined from the entire spectrum) by $\geq$0.1~dex. The discrepancy between [Fe/H] and [M/H] measurements is not fully understood yet. For security, we exclude stars that exhibit this discrepancy, which 
are mostly metal-rich ($\rm [Fe/H]>0.1$) and cool ($T_{\rm eff}<4000$~K). 
This selection has the effect of making our result regarding stars at the high-metallicity end of the distribution {\it less} significant than it would be otherwise, due to the lower counts. 
} 

To ensure reliable measurements of stellar properties, we select stars with spectral signal-to-noise ratio (SNR) above 60 and without warning flags set in the data quality and parameter determination bitmasks\footnote{\url{https://www.sdss.org/dr16/algorithms/bitmasks/\#ListofBitmasks}}.   Specifically, we select stars with $\rm EXTRATARG==0$ and with the 1st, 4th, 9th, 16th and 17th bits of STARFLAG set equal to 0, corresponding respectively to COMMISSIONING, LOW\_SNR, PERSIST\_HIGH, SUSPECT\_RV\_COMBINATION, and SUSPECT\_BROAD\_LINES.  We also require that
the 19th and 23th bits of ASPCAPFLAG be set to 0, corresponding to METALS\_BAD and STAR\_BAD. 
The chemical abundance determinations from the APOGEE pipeline tend to become less robust towards lower effective temperature ($T_{\rm eff}$). Therefore we further exclude stars with $T_{\rm eff}<3500$K \citep{hasselquist2019}. 
We {tested other minimum effective temperature limits, e.g., 4000~K, and confirmed that our results are robust against the choice of this limit.}   

The Galactic bar is an elongated structure that extends out to $5\;$kpc (or Galactic longitude $l\sim25\degree$) at an angle to the line-of-sight to the Galactic center of $\sim25\degree$ \citep[e.g.,][]{stanek1997,wegg2013,wegg2015}.   
In this work, we select stars ``on'' and ``off'' the bar 
{following the definition in \citet{bovy2019}. 
The bar is defined as an ellipse in the Galactic $X-Y$ plane with {a half-length of 5~kpc} and an axis ratio of 0.4, angled 25$^\circ$ from the line of sight to the Galactic Center. The off-bar component is defined as the region outside of the bar ellipse with Galactocentric radius within 5~kpc.} 
Figure~\ref{xy} shows the spatial distribution of APOGEE stars in the $X-Y$ plane. The larger black circle, with a radius of 5$\;$kpc, indicates the region for the selection of the entire sample, while the smaller circle with radius $3\;$kpc is used to separate this inner Galaxy sample into two radial bins. The orange ellipse indicates the bar's spatial definition from \citet{bovy2019}. 

\begin{figure*}
	\centering
	\includegraphics[width=16cm]{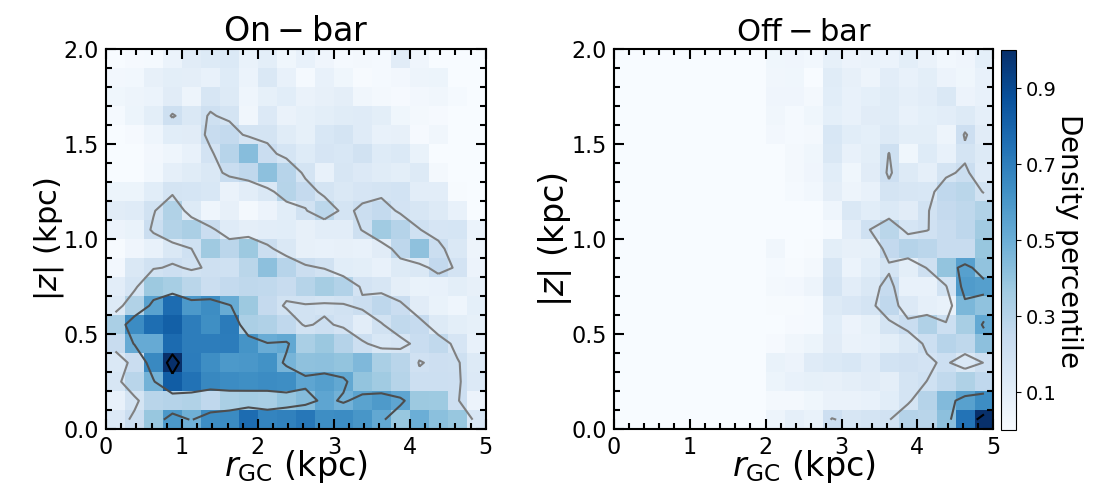}
	\caption{Spatial distribution of on-bar (left-hand panel) and off-bar (right-hand panel) samples in the $r_{\rm GC}$-$|z|$ plane.}
	\label{rz}
\end{figure*}

Figure~\ref{rz} shows the spatial distribution of selected on- and off-bar samples in the $r_{\rm GC}-|z|$ plane.  
It is clear that the on- and off-bar samples 
have very different radial distributions, with peak densities around $r_{\rm GC} \sim 1$~kpc and 5~kpc, respectively. This difference is largely due to the bar definition adopted in this work, which comprises most of the inner $r_{\rm GC} < 3$~kpc region. {The off-bar sample peaks at $r_{\rm GC} \sim 5$~kpc because the off-bar region, by definition, is dominated by the outer annulus whose radial distribution is weighted towards larger $r_{\rm GC}$. The peak density of the on-bar sample at $r_{\rm GC} \sim 1$~kpc is also a result of the bar selection area and observational density distribution, this time weighted towards smaller $r_{\rm GC}$.} {To mitigate potential biases that may be introduced by the different spatial distributions, we perform spatial resampling to ensure the same distribution of the on- and off-bar samples in the $r_{\rm GC}-|z|$ plane (see more details in Section~\ref{sec:resample}).}
In order to study potential spatial variation of stellar populations on/off the bar, we split our on- and off-bar samples into four sub-regions in the $r_{\rm GC}-|z|$ plane.  We use the term ``bulge'' here to refer to the innermost part of the Milky Way, with $r_{\rm GC} < 3$~kpc, to distinguish from the inner disc at slightly larger radii ($r_{\rm GC} = 3-5$~kpc).
\begin{itemize}
\item mid-plane inside the bulge, $r_{\rm GC}<3$~kpc and $|z|<0.5$~kpc (356 stars)
\item off-plane inside the bulge, $r_{\rm GC}<3\;$kpc and $0.8<|z|<1.5\;$kpc (283 stars)
\item mid-plane outside the bulge, $3<r_{\rm GC}<5\;$kpc and $|z|<0.5\;$kpc (2538 stars)
\item off-plane outside the bulge, $3<r_{\rm GC}<5\;$kpc and $0.8<|z|<1.5\;$kpc (3178 stars) 
\end{itemize}

\subsection{Spatial resampling}
\label{sec:resample}

The inner Galaxy has clear radial and vertical gradients in chemical abundances and age (\citealt{zoccali2017,garcia2018,bovy2019,hasselquist2020}). 
As a result, the chemical properties of the on-bar and off-bar regions depend on the radial and vertical distribution of the stars observed in each region. 
To conduct a fair comparison, 
we randomly resample the on-bar stars to have a distribution in the $r_{\rm GC}-|z|$ plane identical to that of the off-bar stars. 
This random resampling is repeated 100 times to obtain the median chemical abundance distribution of the resampled on-bar population.
We note that after the resampling of the on-bar stars to match the radial and vertical distribution of the off-bar stars, the in-bulge on- and off-bar samples (top two in the list above) are concentrated between $r_{\rm GC} \sim 2-3$~kpc and heavily weighted to 3~kpc.

\section{Abundance distribution comparisons} 
\begin{figure*}
	\centering
	\includegraphics[width=16cm,viewport=10 10 1500 830,clip]{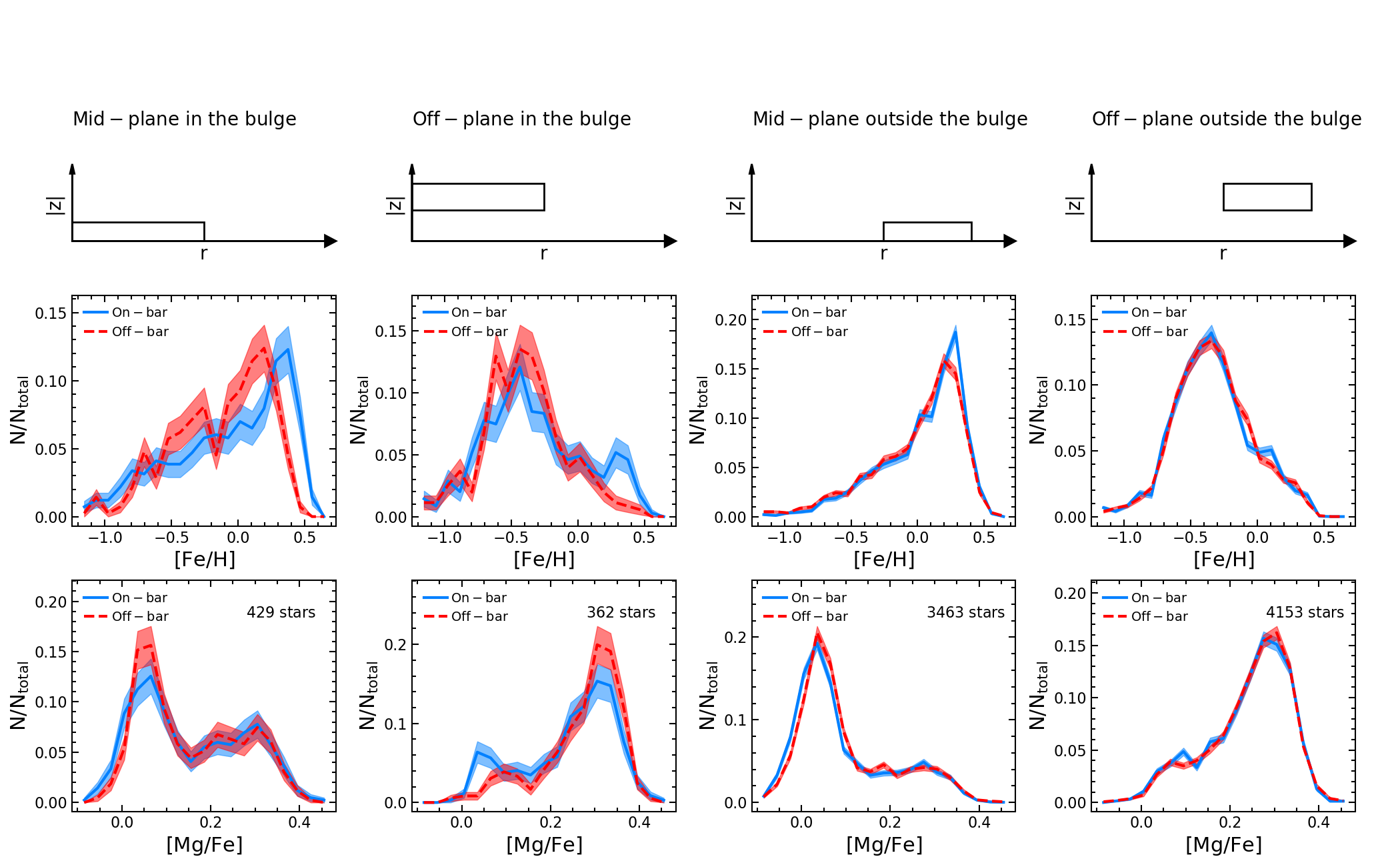}
	\caption{
	[Fe/H] and [Mg/Fe] distributions of stars in different regions.  The top row shows the four regions considered in this work (see Section~\ref{sec:sample}).  The middle and bottom rows contain the distributions of [Fe/H] and [Mg/Fe], respectively, for stars in the region at the top of each column.  On-bar distributions are shown as blue solid lines, and off-bar as red dashed lines. {Shaded areas indicate 1$\sigma$ scatter, assuming Poisson noise.} 
	} 
	\label{mdf-bar}
\end{figure*} 

\subsection{Abundance distribution functions}
We first compare the [Fe/H] and [Mg/Fe] distribution functions (MDF and $\alpha$-DF) of the on-bar and off-bar regions, as illustrated in Figure~\ref{mdf-bar}. 
Each column indicates one of the regions in the $r_{\rm GC}-|z|$ plane described in Section~\ref{sec:sample}. 

In the middle-right and right columns, which show the comparison outside the bulge region, the MDF and $\alpha$-DF of the on-bar and off-bar stars are remarkably consistent, regardless of vertical height. The Galactic bar consists of two main sub-components: a buckled inner bar with a peanut-shaped projected density distribution within $r_{\rm GC}<3\;$kpc, and a planar outer bar that extends to $r_{\rm GC}\sim5\;$kpc and is also called the long or thin bar (e.g., \citealt{wegg2015,barbuy2018}).  
The indistinguishable chemical abundance distributions between the long bar and off-bar disc suggest either very efficient azimuthal mixing between the two structures or that they have experienced rather similar star formation and chemical enrichment histories. The observed azimuthal abundance variation inside the bulge (shown below) and in external barred galaxies \citep{neumann2020}, however, disfavors the azimuthal mixing scenario.

In the left and middle-left panels of Fig.~\ref{mdf-bar}, which show the comparison inside the bulge region, notable differences exist between the MDF and $\alpha$-DF of the on-bar and off-bar stars. {The shaded regions indicate the Poisson noise at a given [Fe/H] and [Mg/Fe].} 
The MDF of the bar in the mid-plane (upper left panel) extends to a higher [Fe/H] than the off-bar disc by $\sim0.2\;$dex, resulting in an excess of very metal-rich stars ($\rm [Fe/H]\sim0.4$) in the $r_{\rm GC} \sim 2-3$~kpc region spanned by our bar stars. The corresponding $\alpha$-DF in the bottom-left panel shows {no significant difference in the plane. However, off the plane (middle-left column) there is a clear excess of metal-rich, low-$\alpha$ stars (${\rm [Fe/H]>0.2\ and\ [Mg/Fe]<0.1}$) on the bar compared to the off-bar sample.} 
To explore the potential effect of stellar parameters (i.e., log($g$) and $T_{\rm eff}$) on the on/off-bar comparison, we resample the stellar parameter distribution of the on-bar sample to be identical to that of the off-bar sample. With this further resampling, the differences in MDF and $\alpha$-DF described above persist, suggesting that they are not caused by differences in stellar parameters of the samples. {Note that the usage of [Fe/H] in this work (instead of [M/H]) excludes a minor fraction of metal-rich stars, which are preferentially located in the bar. Due to the resulting lower counts, then, we expect the measured significance of the excess of metal-rich on-bar stars to be {\it less} than it would be otherwise. 
We also have tested and confirmed that the presence of differences inside the bulge and absence of such differences outside the bulge is not due to the disparate number of bulge and out-of-bulge stars.}

The excess of low-$\alpha$ stars at larger $|z|$ implies that 
these stars are more widely distributed in the vertical direction (i.e., have a larger scale height) in the inner bar than their counterparts in the off-bar disc. This is broadly consistent with previous findings that the inner bar is generally thicker than {the disc outside the bulge} (e.g., \citealt{wegg2013}), which is believed to be a result of dynamical heating through the bar buckling process. The relative deficiency of high-$\alpha$ stars at larger height in the bar is further evidence that these stars, compared to the low-$\alpha$ stars, are less subject to be buckled. {In addition, the lack of azimuthal abundance variation at $3<r_{\rm GC}<5$~kpc suggests} no chemical signature of the buckling process in the long bar outside the bulge. 
It is interesting to note that the presence of considerable differences in the MDF and $\alpha$-DF between the on-bar and off-bar disc within $r_{\rm GC}<3\;$kpc implies {that if there is movement of stars between these regions, the process is either one-directional or inefficient.} 

\subsection{[Mg/Fe]--[Fe/H] distribution}
\begin{figure*}
	\centering
	\includegraphics[width=16cm,viewport=10 10 1500 820,clip]{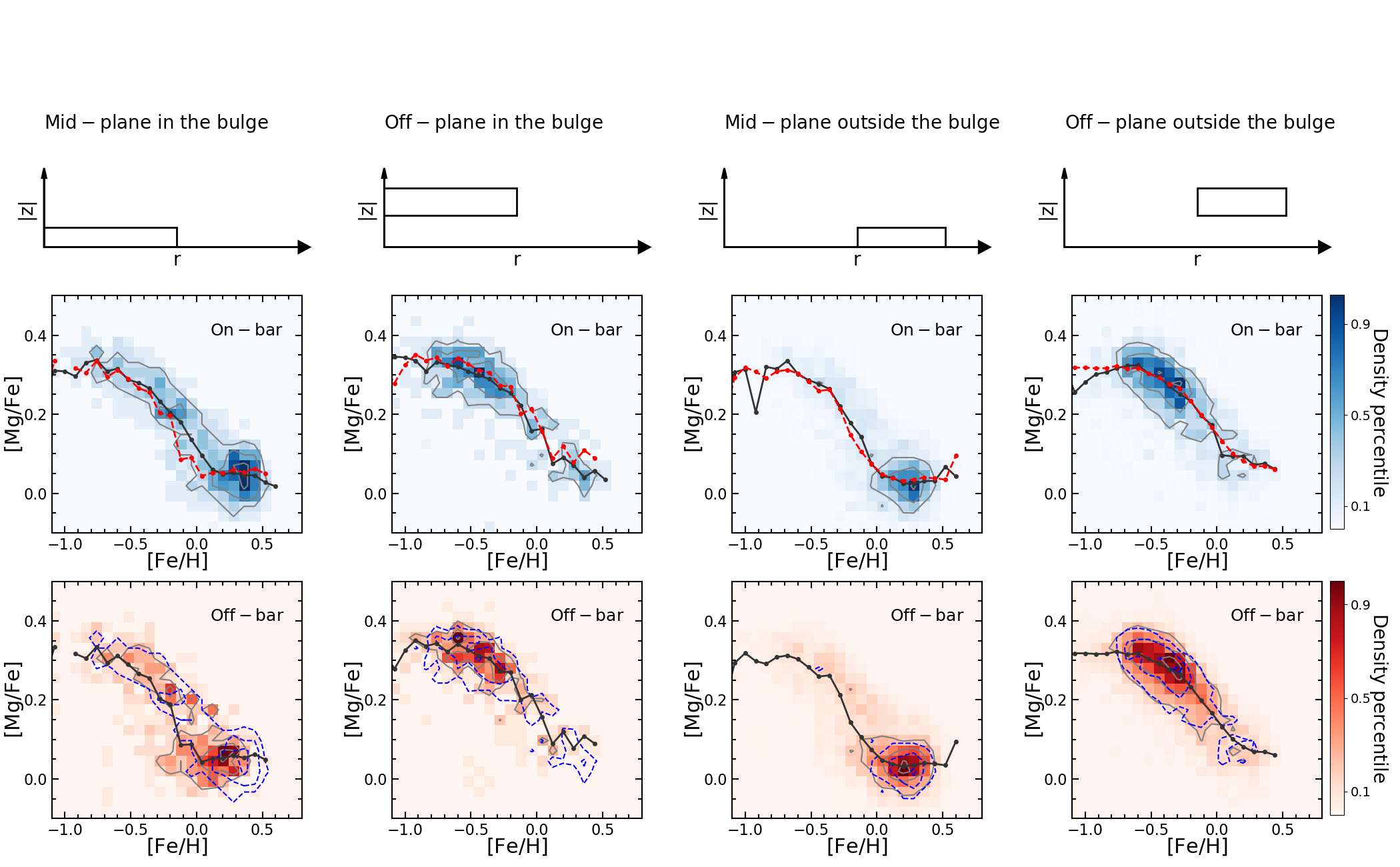}
	\caption{[Mg/Fe]--[Fe/H] distribution of on- (middle row) and off-bar (bottom-row) populations in four regions in the inner Galaxy as indicated in the top row (similar to Figure~\ref{mdf-bar}). The density in each panel is outlined by a grey contour, and black solid lines show the median [MgFe]--[Fe/H] relation. The distribution of the on-bar stars (middle row) is repeated in the bottom row of each column as blue dashed contours. The median [Mg/Fe]--[Fe/H] relation of the off-bar population (bottom row) is repeated in the middle row as red dashed curves. } 
	\label{afe-feh}
\end{figure*} 

\begin{figure*}
	\centering
	\includegraphics[width=16cm]{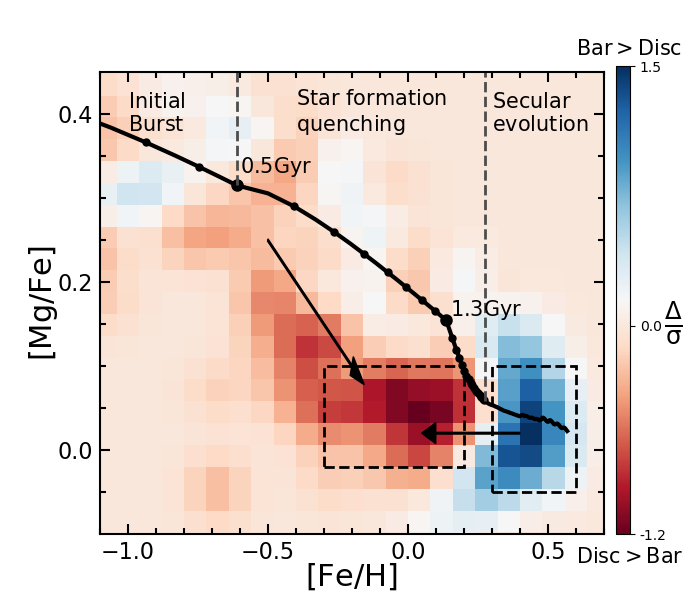}
	\caption{Difference in normalized density ($\Delta$) between the bar and off-bar disc in the plane with $r_{\rm GC} \sim 2-3$~kpc, divided by Poisson noise ($\sigma$). Blue indicates compositions more common of stars in the bar, while red indicates compositions more typical of stars in the off-bar disc. The black line denotes the track of the best-fitted chemical evolution model for the bulge in \citet{lian2020c} (Section~\ref{sec:quenching_accretion}). Small solid circles on the track indicate constant time interval of $0.1\;$Gyr for the first $3\;$Gyr, and the two enlarged circles mark important transition points in the model. Black dashed boxes highlight the dominant differences between the [Mg/Fe]--[Fe/H] distributions of the on-bar and off-bar stars. Diagonal and horizontal arrows illustrate two possible variants of the off-bar disc's evolution track given a SFH with either faster quenching or recent gas accretion, respectively (Section~\ref{sec:quenching_accretion}).} 
	\label{diff}
\end{figure*}

Comparisons of one-dimensional MDFs and $\alpha$-DFs are informative but could hide additional features due to projection effects. 
Thus we unfold the comparison to the [Mg/Fe]--[Fe/H] plane.  
Figure~\ref{afe-feh} shows the density distribution of on-bar (middle row) and off-bar (bottom row) stars in [Mg/Fe]--[Fe/H], with one column for each region in the $r_{\rm GC}-|z|$ plane as indicated in the top row. 
The black solid lines depict the median [Mg/Fe]--[Fe/H] relation in each panel.  
To aid the comparison, we reproduce the [Mg/Fe]--[Fe/H] density distribution of the on-bar stars (middle row) as blue dashed contours in the bottom row. Similarly, we copy the median [Mg/Fe]--[Fe/H] relation of the off-bar population (bottom row) as red dashed lines in the middle row.
We use both the median [Mg/Fe]--[Fe/H] relation and [Mg/Fe]--[Fe/H] contours to aid comparison in distribution shape and density at the same time.  {(Figure~\ref{diff}, described in \S\ref{sec:significance_differences}, shows more directly the differences in the left-hand column.)}

The inter-region comparisons in the [Mg/Fe]--[Fe/H] plane are generally consistent with those from the one-dimensional MDFs and $\alpha$-DFs. 
Outside the bulge, the on-bar and off-bar stars show remarkably consistent [Mg/Fe]--[Fe/H] and density distributions in the [Mg/Fe]--[Fe/H] plane. Inside the bulge, the differences present in the MDF and $\alpha$-DF are visible in the [Mg/Fe]--[Fe/H] distribution plane, as shown in the left and middle-left panels of Figure~\ref{afe-feh}. 

However, there is a feature in the left column that is challenging to identify in the one-dimensional distribution functions. 
The off-bar disc tends to contain fractionally more stars with solar-like [Fe/H] and [Mg/Fe] than the bar. 
Although a similar trend seems present in the MDF and $\alpha$-DF in Fig.~\ref{mdf-bar}, based on abundance distribution functions alone it would likely be considered an effect of the excess of very metal-rich stars and larger scale height of low-$\alpha$ stars in the bar. 
However, the distribution in the [Mg/Fe]--[Fe/H] diagram shows that the relative excess of solar-like abundance stars in the off-bar disc deviates from the [Mg/Fe]--[Fe/H] relation in the bar and results in a different structure of the low-$\alpha$ branch, compared to the on-bar stars. This supports the idea that the relative excess of solar-like abundance stars in the off-bar disc is not caused by a deficiency at other [Fe/H] and [$\alpha$/Fe], but is an independent feature (Section~\ref{sec:significance_differences}).  

 

\section{Discussion} 

\subsection{Comparison with other works}
In a recent closely-related work, \citet{bovy2019} conducted {one of the first} comparisons of the stellar chemical abundance distributions between on-bar and off-bar stars using data from APOGEE DR16.
The comparison was performed 
for the bar and disc within $r_{\rm GC}<5$~kpc and $|z| < 0.3$~kpc (without spatial resampling). The authors found that the bar and the off-bar disc have rather similar distributions in the [Mg/Fe]--[Fe/H] plane, except for a clear excess of old, metal-poor stars in the bar that was interpreted as evidence of an early bar formation. 
In contrast, in a study focused on the on-bar/off-bar comparison at $r_{\rm GC} \sim 3-4$~kpc, \citet{wegg2019} found the in-bar stars (defined spatially and kinematically) to be on average more metal-rich than the surrounding disc stars at similar radii. {\citet{queiroz2020b} also find that the most metal-rich stars are trapped in the bar at $r_{\rm GC}<5$~kpc.}

In this study, we conduct the on-bar and off-bar disc comparison within four spatial bins. To account for the metallicity gradients, the on-bar stars are resampled to have a distribution in radius and height identical to the off-bar sample. Thus we are looking at strictly azimuthal variations at fixed radius and height. 
The near-identical chemical abundance distribution between the bar and disc shown in \citet{bovy2019} is confirmed outside the bulge region ($3 < r_{\rm GC} < 5$~kpc). However, our comparison reveals a difference inside the bulge ($r_{\rm GC} \sim 2-3$~kpc), with relatively more very metal-rich stars and fewer solar-metallicity stars in the bar than in the disc. 
These differences suggest that the buckled inner bar has experienced a somewhat different SFH from the disc at the same Galactocentric radius (see Section~\ref{sec:significance_differences}).

The main reason for the different conclusions regarding the bar's chemistry in \citet{bovy2019} and this work appears to be that the comparisons in these two works are conducted on samples with different spatial distributions. The inner Galaxy has a clear radial metallicity gradient, with the populations at smaller radii being more metal-poor on average \citep[][confirmed in our data before resampling]{bovy2019}. This positive metallicity gradient, combined with the concentration of the raw on-bar sample at small $r_{\rm GC}$, gives rise to a bar that is {\it on average} more metal-poor than the off-bar disc within the larger bar radius of 5~kpc. 


It is very interesting to note that nearby barred galaxies exhibit intriguing abundance differences within their inner regions, with the elongated bar being slightly more metal-rich and $\alpha$-poor than the surrounding disc \citep{neumann2020}. Similarly metal-enhanced bars, compared to the disc, are also seen in barred galaxies in cosmological simulations \citep{buck2018,fragkoudi2020}. {In particular, in the simulation by \citet{buck2018}, the MDF of the bar also extends to higher metallicity than the disc.}
{These differences} in the stellar compositions between the bar and disc is {qualitatively} consistent with the Milky Way stellar observations as presented in this study.  This implies that the bars in our Galaxy and in external galaxies likely share a common formation history.

\subsection{Significance of the differences and possible explanations}
\label{sec:significance_differences}

In this work we find considerable differences between the chemical abundance distribution of the {bar and off-bar disc} stars at $r_{\rm GC}\sim2-3\;$kpc and $|z|<0.5$~kpc. To highlight these differences, we take the difference of the normalized [Mg/Fe]--[Fe/H] density distributions of these two components ($\Delta=$~bar--disc). Assuming Poisson noise (i.e., square root of the number of stars, $\sigma$), we estimate the significance of the difference ($\Delta/\sigma$) as shown in Figure~\ref{diff}.
Blue shading indicates chemical compositions that are relatively more common in the bar, and red shading highlights chemical compositions more common in the off-bar disc. 
The relative excesses of metal-rich stars in the bar and solar-like abundance stars in the disc can be clearly seen. 

Although the significance in each individual pixel is generally smaller than 2, the total significance for these two features (enclosed by black dashed boxes) is much higher: 5.3 for the red blob at $\rm ([Fe/H],[Mg/Fe])=(-0.1,+0.1)$ and 5.4 for the blue one at $\rm ([Fe/H],[Mg/Fe])=(+0.5,+0.05)$. {We also calculate the significance of these two features by performing a bootstrap resampling. We resample the off-bar sample 100 times (the same as the resampling of on-bar stars as described in \textsection2.2) and calculate the difference between the bar and off-bar disc in the [$\alpha$/Fe]--[Fe/H] plane for each resampling. The significance in a given [$\alpha$/Fe]--[Fe/H] bin is estimated as the average value of the 100 resamplings divided by the standard deviation. With this method, the significance of the red and blue blobs are 4.3 and 6.3, respectively, which confirm that these on-bar/off-bar differences are statistically significant.}

These abundance differences between the bar and disc {reflect differences in the 
physical processes that regulate their chemical evolution, including radial migration, gas outflow, and SFH. 
Under the hypothesis that radial migration has shaped the observed abundance differences, the likely explanation of the excess of solar-like abundance stars in the disc at $r_{\rm GC}<3\;$kpc would be that inward migration of these stars is more effective onto the disc than onto the bar. This selective radial migration behavior,    
however, lacks observational support and conflicts with simulations in which inwardly migrating stars are mostly captured by the bar \citep{halle2015}. 
Strong outflows, which could suppress the formation of metal-rich stars, are also unlikely to be responsible for the abundance differences, because this scenario would 
require finely tuned differences between in the star formation-driven outflow in the central part of the Milky Way, with stronger outflow in the disc than in the adjacent bar. 
Therefore we consider a varying SFH a more likely explanation for the observed abundance differences. 
In the next section, we explore two possibilities for a varying SFH. }

\subsection{Effects of quenching and gas accretion}
\label{sec:quenching_accretion}

In \citet{lian2020c} we proposed a three-phase SFH for the inner Galaxy within $r_{\rm GC}<3\;$kpc (not divided into on-bar and off-bar regions). That history consists of an initial star burst, followed by a rapid star-formation quenching epoch, and then a long-term secular phase of low-level star formation. 
The black curve in Fig.~\ref{diff} shows the chemical evolution track corresponding to the best-fitted model of \citet{lian2020c}. Solid black circles on the track indicate constant time intervals of $0.1\;$Gyr for the first $3\;$Gyr after the initial star burst. {See a more detailed comparison in that paper between this model and other inner galaxy chemical evolution models \citep[e.g.,][]{matteucci2019}.}

The two enlarged circles at 0.5 and $1.3\;$Gyr highlight two important transition points in the model: when the gas accretion is switched off, and when star formation efficiency starts to decline. The cessation of gas accretion marks the end of the initial star burst and the onset of a decrease in star formation rate (SFR). The drop in star formation efficiency speeds up the decrease in SFR, which results in a density gap along the population's [Mg/Fe]--[Fe/H] track and a [Mg/Fe] offset between the high- and low-$\alpha$ branches. 
Under this three-phase SFH, the [Mg/Fe]-[Fe/H] diagram can be split into three sections, corresponding to the chemical evolution during the three phases, as indicated by the vertical dashed lines in Fig.~\ref{diff}. 

Based on this star formation framework, we propose two viable scenarios to explain the observed fractional excesses of metal-rich stars in the inner bar and of solar-like abundance stars in the corresponding off-bar disc. 

\subsubsection{Faster quenching in the disc?}
\label{sec:faster_quenching}
The first possible scenario is that the off-bar disc experienced a faster quenching process than the bar, which results in a more rapid decrease in SFR and thus in [Mg/Fe], {given less enrichment in [Fe/H]} during the star formation quenching phase. As a result, the chemical evolution track of the off-bar disc during the quenching phase will be steeper than the fiducial model, as illustrated by the diagonal arrow in Fig.~\ref{diff}. The disc will thus start its secular phase at lower [Fe/H] (around solar abundance). 

Due to reduced star formation during the quenching phase, the metal enrichment in the disc, from this point, is delayed compared to the bar. In this way, the disc will form more solar-like abundance stars than the bar but fewer very metal-rich stars, as observed. 
{The underlying physical mechanism responsible for this faster quenching scenario in the disc is unclear. One possibility is that the rotating bar removes gas from the inner disc and pushes it along the leading edge of the bar towards the Galactic center. As a result, star formation is suppressed in the disc but continues on the leading edge of the bar \citep{neumann2019,Fraser-McKelvie2020}. This bar-induced star formation quenching could also qualitatively explain the quenched SFH in the off-bar inner disc observed in external galaxies \citep{james2016,james2018} and in simulations \citep{Donohoe-Keyes2019}, as well as the azimuthal variation of chemical abundances and SFH in external barred galaxies \citep{neumann2020}.  
}

\subsubsection{Additional gas accretion in the disc?}
\label{sec:more_accretion}
Alternatively, the off-bar disc may share the same early SFH as the bar, but have accreted metal-poor gas at some point during the secular evolution phase. This newly accreted metal-poor gas dilutes the total metal abundance in the disc and draws the evolution track backward (to lower metallicity) in the [Mg/Fe]--[Fe/H] plane, as indicated by the horizontal arrow in the bottom-right corner in Fig.~\ref{diff}. 
{This metal-poor gas accretion hinders the formation of metal-rich stars and instead boosts the number of stars formed at lower metallicities in the disc.}
Such late metal-poor gas accretion has been suggested in many recent works to explain the [$\alpha$/Fe]--[Fe/H] pattern of low-$\alpha$ branch in the {disc outside the bulge} \citep[e.g.,][]{haywood2018,spitoni2019,lian2020,lian2020b}. 

\subsubsection{Disentangling the two scenarios}
So far we lack solid evidence from observations or simulations to disentangle these two different SFH solutions. 
Extragalactic observations reveal a complex picture of the connection between the presence of a bar and star formation/gas content in the host galaxy. For example, while many works find an increased bar fraction in galaxies with lower star formation activity \citep{masters2011} and gas fraction \citep{masters2012,newnham2019}, other studies find enhanced central star formation \citep{wang2012,lin2017} and high gas concentration in barred galaxies \citep{chown2019}. 
Hydrodynamical simulations predict gas inflow along the bar towards the galactic center within the bar cororation radius \citep{athanassoula2013,spinoso2017}. This seems to disfavor the inhomogeneous gas accretion scenario described in Section~\ref{sec:more_accretion}, in which metal-poor gas inflow preferentially ends up in the off-bar disc.  

Note that the faster quenching scenario implies the bar is already present at early times ($\sim1\;$Gyr after initial star formation), while the inhomogeneous gas accretion scenario has a looser requirement on the bar formation epoch. 
Constraints from simulations to distinguish these scenarios are scant, however.  The formation time of a bar in simulated galaxies varies dramatically, {from less than 1~Gyr to several Gyr after the formation of the disc, and depends} heavily on host galaxy properties and environment, such as disc gas fraction, halo structure, and satellite accretion history \citep{athanassoula2013,spinoso2017,zana2018}. 

Although very difficult to measure, ages of stars in the inner MW provide critical---perhaps the  best---constraints on the SFH of our bar and disc and is therefore a promising approach to differentiate the two viable scenarios. 
Given the uninterrupted chemical enrichment in the ``disc quenched faster'' scenario,  a positive age-metallicity relation of the low-$\alpha$ population in the off-bar disc is expected. In contrast, the inhomogenous accretion scenario would predict a complicated age-metallicity relation, with a multimodal age distribution at many metallicities. 
\citet{hasselquist2020} will provide robust age measurements for a large sample of APOGEE bulge stars. Despite the uncertainty in the age-metallicity relation due to the 0.2--0.3~dex individual age uncertainties, preliminary assessment of these ages suggests a single age sequence in the low-$\alpha$ population in the disc, favoring the faster quenching scenario.

\section{Summary}
In this work we investigate the star formation history of multiple regions in the inner Galaxy: the on-bar and off-bar disc, further divided by Galactic radius and height. We explore these histories by analyzing the [Mg/Fe] and [Fe/H] distributions of stars in each component using abundances derived by APOGEE. 
To avoid potential biases in the MDF and $\alpha$-DF introduced by radial and vertical chemical abundance gradients in the inner Galaxy, we resample the on-bar stars to achieve the same spatial distribution in the $r_{\rm GC}-|z|$ plane as the off-bar sample. The on- and off-bar samples are then split into four sub-regions in the $r_{\rm GC}-|z|$ plane. 

By comparing the one-dimensional MDF, one-dimensional $\alpha$-DF, and two-dimensional [Mg/Fe]--[Fe/H] distributions between the on-bar and off-bar stars in each region, we find the bar and disc outside the bulge region ($3<r_{\rm GC}<5\;$kpc) to be remarkably consistent; however, those inside the bulge ($r_{\rm GC}<3\;$kpc, predominately at 2--3~kpc) show clear differences. 

The first of these results suggests the long bar and disc share a common SFH. In contrast, the on-bar region in the plane at $r_{\rm GC} < 3$~kpc contains significantly more very metal-rich stars ([Fe/H]$\sim0.4$), but fewer solar-like abundance stars, compared to the off-bar disc at the same radius. This difference implies an azimuthally varying SFH in the inner Galaxy without efficient azimuthal mixing. We also find the low-$\alpha$ population in the bar {tends to have a wider vertical distribution} than its counterpart in the off-bar disc, which {might} be a chemical signature of vertical heating through a bar buckling process {(see \textsection 3.1 for more discussion)}. The absence of this feature outside the bulge region implies the long bar has not yet buckled. 


In a companion paper \citep{lian2020c}, we propose a three-phase SFH for the integrated Galactic bulge that consists of an initial star burst, then a rapid star formation quenching, and finally a long-term secular evolution phase. 
Under this three-phase SFH, the observed differences in the abundance distributions between the bar and disc could be attributed to minor differences in SFH. One possibility is that the off-bar disc experienced a faster early star formation quenching; another possibility is that it instead underwent a recent metal-poor gas accretion. 
Both scenarios could in principle explain the 
higher ratio of super-solar to solar-metallicity stars in the bar, compared to the off-bar disc.
Existing observations---in the MW, in extragalactic the best of systems, or in simulations---are not adequate to  disentangle these two scenarios. Given early results from stellar age measurements and gas kinematics in simulations, the faster quenching scenario is slightly favored.  

\section*{Acknowledgements}
We are grateful to the anonymous referee for useful comments which significantly improve the clarity of the paper. S.H. is supported by an NSF Astronomy and Astrophysics Postdoctoral Fellowship under award AST-1801940. The Science, Technology and Facilities Council is acknowledged by JN for support through the Consolidated Grant Cosmology and Astrophysics at Portsmouth, ST/S000550/1.

Funding for the Sloan Digital Sky Survey IV has been provided by the Alfred P. Sloan Foundation, the U.S. Department of Energy Office of Science, and the Participating Institutions. SDSS-IV acknowledges
support and resources from the Center for High-Performance Computing at
the University of Utah. The SDSS web site is www.sdss.org.

SDSS-IV is managed by the Astrophysical Research Consortium for the 
Participating Institutions of the SDSS Collaboration including the 
Brazilian Participation Group, the Carnegie Institution for Science, 
Carnegie Mellon University, the Chilean Participation Group, the French Participation Group, Harvard-Smithsonian Center for Astrophysics, 
Instituto de Astrof\'isica de Canarias, The Johns Hopkins University, Kavli Institute for the Physics and Mathematics of the Universe (IPMU) / 
University of Tokyo, the Korean Participation Group, Lawrence Berkeley National Laboratory, 
Leibniz Institut f\"ur Astrophysik Potsdam (AIP),  
Max-Planck-Institut f\"ur Astronomie (MPIA Heidelberg), 
Max-Planck-Institut f\"ur Astrophysik (MPA Garching), 
Max-Planck-Institut f\"ur Extraterrestrische Physik (MPE), 
National Astronomical Observatories of China, New Mexico State University, 
New York University, University of Notre Dame, 
Observat\'ario Nacional / MCTI, The Ohio State University, 
Pennsylvania State University, Shanghai Astronomical Observatory, 
United Kingdom Participation Group,
Universidad Nacional Aut\'onoma de M\'exico, University of Arizona, 
University of Colorado Boulder, University of Oxford, University of Portsmouth, 
University of Utah, University of Virginia, University of Washington, University of Wisconsin, 
Vanderbilt University, and Yale University.

\section*{Data Availability}
The data underlying this article is from an internal incremental release of the SDSS-IV/APOGEE survey, following the SDSS-IV public Data Release 16 (using reduction pipeline version r13). This incremental release is not publicly available now but will be included in the final public data release of SDSS-IV.   
The chemical evolution model results are available upon request. 


\bibliographystyle{mnras}
\bibliography{Jianhui}{}

\begin{thebibliography}{}
\makeatletter
\relax
\def\mn@urlcharsother{\let\do\@makeother \do\$\do\&\do\#\do\^\do\_\do\%\do\~}
\def\mn@doi{\begingroup\mn@urlcharsother \@ifnextchar [ {\mn@doi@}
  {\mn@doi@[]}}
\def\mn@doi@[#1]#2{\def\@tempa{#1}\ifx\@tempa\@empty \href
  {http://dx.doi.org/#2} {doi:#2}\else \href {http://dx.doi.org/#2} {#1}\fi
  \endgroup}
\def\mn@eprint#1#2{\mn@eprint@#1:#2::\@nil}
\def\mn@eprint@arXiv#1{\href {http://arxiv.org/abs/#1} {{\tt arXiv:#1}}}
\def\mn@eprint@dblp#1{\href {http://dblp.uni-trier.de/rec/bibtex/#1.xml}
  {dblp:#1}}
\def\mn@eprint@#1:#2:#3:#4\@nil{\def\@tempa {#1}\def\@tempb {#2}\def\@tempc
  {#3}\ifx \@tempc \@empty \let \@tempc \@tempb \let \@tempb \@tempa \fi \ifx
  \@tempb \@empty \def\@tempb {arXiv}\fi \@ifundefined
  {mn@eprint@\@tempb}{\@tempb:\@tempc}{\expandafter \expandafter \csname
  mn@eprint@\@tempb\endcsname \expandafter{\@tempc}}}

\bibitem[\protect\citeauthoryear{{Aguerri}, {M{\'e}ndez-Abreu}  \&
  {Corsini}}{{Aguerri} et~al.}{2009}]{aguerri2009}
{Aguerri} J.~A.~L.,  {M{\'e}ndez-Abreu} J.,   {Corsini} E.~M.,  2009, \mn@doi
  [\aap] {10.1051/0004-6361:200810931}, \href
  {https://ui.adsabs.harvard.edu/abs/2009A&A...495..491A} {495, 491}

\bibitem[\protect\citeauthoryear{{Ahumada} et~al.,}{{Ahumada}
  et~al.}{2019}]{ahumada2020}
{Ahumada} R.,  et~al., 2019, arXiv e-prints, \href
  {https://ui.adsabs.harvard.edu/abs/2019arXiv191202905A} {p. arXiv:1912.02905}

\bibitem[\protect\citeauthoryear{{Athanassoula}, {Machado}  \&
  {Rodionov}}{{Athanassoula} et~al.}{2013}]{athanassoula2013}
{Athanassoula} E.,  {Machado} R. E.~G.,   {Rodionov} S.~A.,  2013, \mn@doi
  [\mnras] {10.1093/mnras/sts452}, \href
  {https://ui.adsabs.harvard.edu/abs/2013MNRAS.429.1949A} {429, 1949}

\bibitem[\protect\citeauthoryear{{Babusiaux} et~al.,}{{Babusiaux}
  et~al.}{2010}]{babusiaux2010}
{Babusiaux} C.,  et~al., 2010, \mn@doi [\aap] {10.1051/0004-6361/201014353},
  \href {https://ui.adsabs.harvard.edu/abs/2010A&A...519A..77B} {519, A77}

\bibitem[\protect\citeauthoryear{{Barbuy}, {Chiappini}  \& {Gerhard}}{{Barbuy}
  et~al.}{2018}]{barbuy2018}
{Barbuy} B.,  {Chiappini} C.,   {Gerhard} O.,  2018, \mn@doi [\araa]
  {10.1146/annurev-astro-081817-051826}, \href
  {https://ui.adsabs.harvard.edu/abs/2018ARA&A..56..223B} {56, 223}

\bibitem[\protect\citeauthoryear{{Bensby} et~al.,}{{Bensby}
  et~al.}{2013}]{bensby2013}
{Bensby} T.,  et~al., 2013, \mn@doi [\aap] {10.1051/0004-6361/201220678}, \href
  {https://ui.adsabs.harvard.edu/abs/2013A&A...549A.147B} {549, A147}

\bibitem[\protect\citeauthoryear{{Bensby} et~al.,}{{Bensby}
  et~al.}{2017}]{bensby2017}
{Bensby} T.,  et~al., 2017, \mn@doi [\aap] {10.1051/0004-6361/201730560}, \href
  {https://ui.adsabs.harvard.edu/abs/2017A&A...605A..89B} {605, A89}

\bibitem[\protect\citeauthoryear{{Binney}, {Gerhard}, {Stark}, {Bally}  \&
  {Uchida}}{{Binney} et~al.}{1991}]{binney1991}
{Binney} J.,  {Gerhard} O.~E.,  {Stark} A.~A.,  {Bally} J.,   {Uchida} K.~I.,
  1991, \mn@doi [\mnras] {10.1093/mnras/252.2.210}, \href
  {https://ui.adsabs.harvard.edu/abs/1991MNRAS.252..210B} {252, 210}

\bibitem[\protect\citeauthoryear{{Blanton} et~al.,}{{Blanton}
  et~al.}{2017}]{blanton2017}
{Blanton} M.~R.,  et~al., 2017, \mn@doi [\aj] {10.3847/1538-3881/aa7567}, \href
  {https://ui.adsabs.harvard.edu/abs/2017AJ....154...28B} {154, 28}

\bibitem[\protect\citeauthoryear{{Blitz} \& {Spergel}}{{Blitz} \&
  {Spergel}}{1991}]{blitz1991}
{Blitz} L.,  {Spergel} D.~N.,  1991, \mn@doi [\apj] {10.1086/170535}, \href
  {https://ui.adsabs.harvard.edu/abs/1991ApJ...379..631B} {379, 631}

\bibitem[\protect\citeauthoryear{{Bovy}, {Leung}, {Hunt}, {Mackereth},
  {Garc{\'\i}a-Hern{\'a}ndez}  \& {Roman-Lopes}}{{Bovy}
  et~al.}{2019}]{bovy2019}
{Bovy} J.,  {Leung} H.~W.,  {Hunt} J. A.~S.,  {Mackereth} J.~T.,
  {Garc{\'\i}a-Hern{\'a}ndez} D.~A.,   {Roman-Lopes} A.,  2019, \mn@doi
  [\mnras] {10.1093/mnras/stz2891}, \href
  {https://ui.adsabs.harvard.edu/abs/2019MNRAS.490.4740B} {490, 4740}

\bibitem[\protect\citeauthoryear{{Bowen} \& {Vaughan}}{{Bowen} \&
  {Vaughan}}{1973}]{bowen1973}
{Bowen} I.~S.,  {Vaughan} A.~H. J.,  1973, \mn@doi [\ao]
  {10.1364/AO.12.001430}, \href
  {https://ui.adsabs.harvard.edu/abs/1973ApOpt..12.1430B} {12, 1430}

\bibitem[\protect\citeauthoryear{{Buck}, {Ness}, {Macci{\`o}}, {Obreja}  \&
  {Dutton}}{{Buck} et~al.}{2018}]{buck2018}
{Buck} T.,  {Ness} M.~K.,  {Macci{\`o}} A.~V.,  {Obreja} A.,   {Dutton} A.~A.,
  2018, \mn@doi [\apj] {10.3847/1538-4357/aac890}, \href
  {https://ui.adsabs.harvard.edu/abs/2018ApJ...861...88B} {861, 88}

\bibitem[\protect\citeauthoryear{{Buta} et~al.,}{{Buta}
  et~al.}{2015}]{buta2015}
{Buta} R.~J.,  et~al., 2015, \mn@doi [\apjs] {10.1088/0067-0049/217/2/32},
  \href {https://ui.adsabs.harvard.edu/abs/2015ApJS..217...32B} {217, 32}

\bibitem[\protect\citeauthoryear{{Chiappini}, {Matteucci}  \&
  {Gratton}}{{Chiappini} et~al.}{1997}]{chiappini1997}
{Chiappini} C.,  {Matteucci} F.,   {Gratton} R.,  1997, \mn@doi [\apj]
  {10.1086/303726}, \href
  {https://ui.adsabs.harvard.edu/abs/1997ApJ...477..765C} {477, 765}

\bibitem[\protect\citeauthoryear{{Chown} et~al.,}{{Chown}
  et~al.}{2019}]{chown2019}
{Chown} R.,  et~al., 2019, \mn@doi [\mnras] {10.1093/mnras/stz349}, \href
  {https://ui.adsabs.harvard.edu/abs/2019MNRAS.484.5192C} {484, 5192}

\bibitem[\protect\citeauthoryear{{Cunha} \& {Smith}}{{Cunha} \&
  {Smith}}{2006}]{cunha2006}
{Cunha} K.,  {Smith} V.~V.,  2006, \mn@doi [\apj] {10.1086/507673}, \href
  {https://ui.adsabs.harvard.edu/abs/2006ApJ...651..491C} {651, 491}

\bibitem[\protect\citeauthoryear{{Donohoe-Keyes}, {Martig}, {James}  \&
  {Kraljic}}{{Donohoe-Keyes} et~al.}{2019}]{Donohoe-Keyes2019}
{Donohoe-Keyes} C.~E.,  {Martig} M.,  {James} P.~A.,   {Kraljic} K.,  2019,
  \mn@doi [\mnras] {10.1093/mnras/stz2474}, \href
  {https://ui.adsabs.harvard.edu/abs/2019MNRAS.489.4992D} {489, 4992}

\bibitem[\protect\citeauthoryear{{Fragkoudi} et~al.,}{{Fragkoudi}
  et~al.}{2020}]{fragkoudi2020}
{Fragkoudi} F.,  et~al., 2020, \mn@doi [\mnras] {10.1093/mnras/staa1104}, \href
  {https://ui.adsabs.harvard.edu/abs/2020MNRAS.494.5936F} {494, 5936}

\bibitem[\protect\citeauthoryear{{Fraser-McKelvie} et~al.,}{{Fraser-McKelvie}
  et~al.}{2020}]{Fraser-McKelvie2020}
{Fraser-McKelvie} A.,  et~al., 2020, \mn@doi [\mnras] {10.1093/mnras/staa1416},
  \href {https://ui.adsabs.harvard.edu/abs/2020MNRAS.495.4158F} {495, 4158}

\bibitem[\protect\citeauthoryear{{Freeman} et~al.,}{{Freeman}
  et~al.}{2013}]{freeman2013}
{Freeman} K.,  et~al., 2013, \mn@doi [\mnras] {10.1093/mnras/sts305}, \href
  {https://ui.adsabs.harvard.edu/abs/2013MNRAS.428.3660F} {428, 3660}

\bibitem[\protect\citeauthoryear{{Garc{\'\i}a P{\'e}rez} et~al.,}{{Garc{\'\i}a
  P{\'e}rez} et~al.}{2016}]{garcia2016}
{Garc{\'\i}a P{\'e}rez} A.~E.,  et~al., 2016, \mn@doi [\aj]
  {10.3847/0004-6256/151/6/144}, \href
  {https://ui.adsabs.harvard.edu/abs/2016AJ....151..144G} {151, 144}

\bibitem[\protect\citeauthoryear{{Garc{\'\i}a P{\'e}rez} et~al.,}{{Garc{\'\i}a
  P{\'e}rez} et~al.}{2018}]{garcia2018}
{Garc{\'\i}a P{\'e}rez} A.~E.,  et~al., 2018, \mn@doi [\apj]
  {10.3847/1538-4357/aa9d88}, \href
  {https://ui.adsabs.harvard.edu/abs/2018ApJ...852...91G} {852, 91}

\bibitem[\protect\citeauthoryear{{Gunn} et~al.,}{{Gunn}
  et~al.}{2006}]{gunn2006}
{Gunn} J.~E.,  et~al., 2006, \mn@doi [\aj] {10.1086/500975}, \href
  {https://ui.adsabs.harvard.edu/abs/2006AJ....131.2332G} {131, 2332}

\bibitem[\protect\citeauthoryear{{Halle}, {Di Matteo}, {Haywood}  \&
  {Combes}}{{Halle} et~al.}{2015}]{halle2015}
{Halle} A.,  {Di Matteo} P.,  {Haywood} M.,   {Combes} F.,  2015, \mn@doi
  [\aap] {10.1051/0004-6361/201525612}, \href
  {https://ui.adsabs.harvard.edu/abs/2015A&A...578A..58H} {578, A58}

\bibitem[\protect\citeauthoryear{{Hasselquist} et~al.,}{{Hasselquist}
  et~al.}{2019}]{hasselquist2019}
{Hasselquist} S.,  et~al., 2019, \mn@doi [\apj] {10.3847/1538-4357/aaf859},
  \href {https://ui.adsabs.harvard.edu/abs/2019ApJ...871..181H} {871, 181}

\bibitem[\protect\citeauthoryear{{Hasselquist} et~al.,}{{Hasselquist}
  et~al.}{2020}]{hasselquist2020}
{Hasselquist} S.,  et~al., 2020, arXiv e-prints, \href
  {https://ui.adsabs.harvard.edu/abs/2020arXiv200803603H} {p. arXiv:2008.03603}

\bibitem[\protect\citeauthoryear{{Haywood}, {Di Matteo}, {Lehnert}, {Snaith},
  {Fragkoudi}  \& {Khoperskov}}{{Haywood} et~al.}{2018}]{haywood2018}
{Haywood} M.,  {Di Matteo} P.,  {Lehnert} M.,  {Snaith} O.,  {Fragkoudi} F.,
  {Khoperskov} S.,  2018, \mn@doi [\aap] {10.1051/0004-6361/201731363}, \href
  {https://ui.adsabs.harvard.edu/abs/2018A&A...618A..78H} {618, A78}

\bibitem[\protect\citeauthoryear{{Hill} et~al.,}{{Hill}
  et~al.}{2011}]{hill2011}
{Hill} V.,  et~al., 2011, \mn@doi [\aap] {10.1051/0004-6361/200913757}, \href
  {https://ui.adsabs.harvard.edu/abs/2011A&A...534A..80H} {534, A80}

\bibitem[\protect\citeauthoryear{{Holtzman}, {Harrison}  \&
  {Coughlin}}{{Holtzman} et~al.}{2010}]{holtzman2010}
{Holtzman} J.~A.,  {Harrison} T.~E.,   {Coughlin} J.~L.,  2010, \mn@doi
  [Advances in Astronomy] {10.1155/2010/193086}, \href
  {https://ui.adsabs.harvard.edu/abs/2010AdAst2010E..46H} {2010, 193086}

\bibitem[\protect\citeauthoryear{{James} \& {Percival}}{{James} \&
  {Percival}}{2016}]{james2016}
{James} P.~A.,  {Percival} S.~M.,  2016, \mn@doi [\mnras]
  {10.1093/mnras/stv2978}, \href
  {https://ui.adsabs.harvard.edu/abs/2016MNRAS.457..917J} {457, 917}

\bibitem[\protect\citeauthoryear{{James} \& {Percival}}{{James} \&
  {Percival}}{2018}]{james2018}
{James} P.~A.,  {Percival} S.~M.,  2018, \mn@doi [\mnras]
  {10.1093/mnras/stx2990}, \href
  {https://ui.adsabs.harvard.edu/abs/2018MNRAS.474.3101J} {474, 3101}

\bibitem[\protect\citeauthoryear{{Jogee} et~al.,}{{Jogee}
  et~al.}{2004}]{jogee2004}
{Jogee} S.,  et~al., 2004, \mn@doi [\apjl] {10.1086/426138}, \href
  {https://ui.adsabs.harvard.edu/abs/2004ApJ...615L.105J} {615, L105}

\bibitem[\protect\citeauthoryear{{J{\"o}nsson} et~al.,}{{J{\"o}nsson}
  et~al.}{2018}]{jonsson2018}
{J{\"o}nsson} H.,  et~al., 2018, \mn@doi [\aj] {10.3847/1538-3881/aad4f5},
  \href {https://ui.adsabs.harvard.edu/abs/2018AJ....156..126J} {156, 126}

\bibitem[\protect\citeauthoryear{{J{\"o}nsson} et~al.,}{{J{\"o}nsson}
  et~al.}{2020}]{jonsson2020}
{J{\"o}nsson} H.,  et~al., 2020, arXiv e-prints, \href
  {https://ui.adsabs.harvard.edu/abs/2020arXiv200705537J} {p. arXiv:2007.05537}

\bibitem[\protect\citeauthoryear{{Kormendy} \& {Kennicutt}}{{Kormendy} \&
  {Kennicutt}}{2004}]{kormendy2004}
{Kormendy} J.,  {Kennicutt} Robert~C. J.,  2004, \mn@doi [\araa]
  {10.1146/annurev.astro.42.053102.134024}, \href
  {https://ui.adsabs.harvard.edu/abs/2004ARA&A..42..603K} {42, 603}

\bibitem[\protect\citeauthoryear{{Kunder} et~al.,}{{Kunder}
  et~al.}{2012}]{kunder2012}
{Kunder} A.,  et~al., 2012, \mn@doi [\aj] {10.1088/0004-6256/143/3/57}, \href
  {https://ui.adsabs.harvard.edu/abs/2012AJ....143...57K} {143, 57}

\bibitem[\protect\citeauthoryear{{Lian} et~al.,}{{Lian}
  et~al.}{2020a}]{lian2020}
{Lian} J.,  et~al., 2020a, \mn@doi [\mnras] {10.1093/mnras/staa867}, \href
  {https://ui.adsabs.harvard.edu/abs/2020MNRAS.494.2561L} {494, 2561}

\bibitem[\protect\citeauthoryear{{Lian} et~al.,}{{Lian}
  et~al.}{2020b}]{lian2020b}
{Lian} J.,  et~al., 2020b, \mn@doi [\mnras] {10.1093/mnras/staa2078}, \href
  {https://ui.adsabs.harvard.edu/abs/2020MNRAS.497.2371L} {497, 2371}

\bibitem[\protect\citeauthoryear{{Lian} et~al.,}{{Lian}
  et~al.}{2020c}]{lian2020c}
{Lian} J.,  et~al., 2020c, \mn@doi [\mnras] {10.1093/mnras/staa2205}, \href
  {https://ui.adsabs.harvard.edu/abs/2020MNRAS.497.3557L} {497, 3557}

\bibitem[\protect\citeauthoryear{{Lin}, {Li}, {He}, {Xiao}  \& {Wang}}{{Lin}
  et~al.}{2017}]{lin2017}
{Lin} L.,  {Li} C.,  {He} Y.,  {Xiao} T.,   {Wang} E.,  2017, \mn@doi [\apj]
  {10.3847/1538-4357/aa657a}, \href
  {https://ui.adsabs.harvard.edu/abs/2017ApJ...838..105L} {838, 105}

\bibitem[\protect\citeauthoryear{{Majewski} et~al.,}{{Majewski}
  et~al.}{2017}]{majewski2017}
{Majewski} S.~R.,  et~al., 2017, \mn@doi [\aj] {10.3847/1538-3881/aa784d},
  \href {https://ui.adsabs.harvard.edu/abs/2017AJ....154...94M} {154, 94}

\bibitem[\protect\citeauthoryear{{Masters} et~al.,}{{Masters}
  et~al.}{2011}]{masters2011}
{Masters} K.~L.,  et~al., 2011, \mn@doi [\mnras]
  {10.1111/j.1365-2966.2010.17834.x}, \href
  {https://ui.adsabs.harvard.edu/abs/2011MNRAS.411.2026M} {411, 2026}

\bibitem[\protect\citeauthoryear{{Masters} et~al.,}{{Masters}
  et~al.}{2012}]{masters2012}
{Masters} K.~L.,  et~al., 2012, \mn@doi [\mnras]
  {10.1111/j.1365-2966.2012.21377.x}, \href
  {https://ui.adsabs.harvard.edu/abs/2012MNRAS.424.2180M} {424, 2180}

\bibitem[\protect\citeauthoryear{{Matteucci}, {Grisoni}, {Spitoni},
  {Zulianello}, {Rojas-Arriagada}, {Schultheis}  \& {Ryde}}{{Matteucci}
  et~al.}{2019}]{matteucci2019}
{Matteucci} F.,  {Grisoni} V.,  {Spitoni} E.,  {Zulianello} A.,
  {Rojas-Arriagada} A.,  {Schultheis} M.,   {Ryde} N.,  2019, \mn@doi [\mnras]
  {10.1093/mnras/stz1647}, \href
  {https://ui.adsabs.harvard.edu/abs/2019MNRAS.487.5363M} {487, 5363}

\bibitem[\protect\citeauthoryear{{Ness} et~al.,}{{Ness}
  et~al.}{2013}]{ness2013}
{Ness} M.,  et~al., 2013, \mn@doi [\mnras] {10.1093/mnras/sts629}, \href
  {https://ui.adsabs.harvard.edu/abs/2013MNRAS.430..836N} {430, 836}

\bibitem[\protect\citeauthoryear{{Ness} et~al.,}{{Ness}
  et~al.}{2016}]{ness2016a}
{Ness} M.,  et~al., 2016, \mn@doi [\apj] {10.3847/0004-637X/819/1/2}, \href
  {https://ui.adsabs.harvard.edu/abs/2016ApJ...819....2N} {819, 2}

\bibitem[\protect\citeauthoryear{{Neumann} et~al.,}{{Neumann}
  et~al.}{2019}]{neumann2019}
{Neumann} J.,  et~al., 2019, \mn@doi [\aap] {10.1051/0004-6361/201834441},
  \href {https://ui.adsabs.harvard.edu/abs/2019A&A...627A..26N} {627, A26}

\bibitem[\protect\citeauthoryear{{Neumann} et~al.,}{{Neumann}
  et~al.}{2020}]{neumann2020}
{Neumann} J.,  et~al., 2020, \mn@doi [\aap] {10.1051/0004-6361/202037604},
  \href {https://ui.adsabs.harvard.edu/abs/2020A&A...637A..56N} {637, A56}

\bibitem[\protect\citeauthoryear{{Newnham}, {Hess}, {Masters}, {Kruk}, {Penny},
  {Lingard}  \& {Smethurst}}{{Newnham} et~al.}{2020}]{newnham2019}
{Newnham} L.,  {Hess} K.~M.,  {Masters} K.~L.,  {Kruk} S.,  {Penny} S.~J.,
  {Lingard} T.,   {Smethurst} R.~J.,  2020, \mn@doi [\mnras]
  {10.1093/mnras/staa064}, \href
  {https://ui.adsabs.harvard.edu/abs/2020MNRAS.492.4697N} {492, 4697}

\bibitem[\protect\citeauthoryear{{Nidever} et~al.,}{{Nidever}
  et~al.}{2015}]{nidever2015}
{Nidever} D.~L.,  et~al., 2015, \mn@doi [\aj] {10.1088/0004-6256/150/6/173},
  \href {https://ui.adsabs.harvard.edu/abs/2015AJ....150..173N} {150, 173}

\bibitem[\protect\citeauthoryear{{Queiroz} et~al.,}{{Queiroz}
  et~al.}{2018}]{queiroz2018}
{Queiroz} A.~B.~A.,  et~al., 2018, \mn@doi [\mnras] {10.1093/mnras/sty330},
  \href {https://ui.adsabs.harvard.edu/abs/2018MNRAS.476.2556Q} {476, 2556}

\bibitem[\protect\citeauthoryear{{Queiroz} et~al.,}{{Queiroz}
  et~al.}{2020a}]{queiroz2020b}
{Queiroz} A.~B.~A.,  et~al., 2020a, arXiv e-prints, \href
  {https://ui.adsabs.harvard.edu/abs/2020arXiv200712915Q} {p. arXiv:2007.12915}

\bibitem[\protect\citeauthoryear{{Queiroz} et~al.,}{{Queiroz}
  et~al.}{2020b}]{queiroz2020}
{Queiroz} A.~B.~A.,  et~al., 2020b, \mn@doi [\aap]
  {10.1051/0004-6361/201937364}, \href
  {https://ui.adsabs.harvard.edu/abs/2020A&A...638A..76Q} {638, A76}

\bibitem[\protect\citeauthoryear{{Rojas-Arriagada} et~al.,}{{Rojas-Arriagada}
  et~al.}{2017}]{rojas2017}
{Rojas-Arriagada} A.,  et~al., 2017, \mn@doi [\aap]
  {10.1051/0004-6361/201629160}, \href
  {https://ui.adsabs.harvard.edu/abs/2017A&A...601A.140R} {601, A140}

\bibitem[\protect\citeauthoryear{{Rojas-Arriagada}, {Zoccali}, {Schultheis},
  {Recio-Blanco}, {Zasowski}, {Minniti}, {J{\"o}nsson}  \&
  {Cohen}}{{Rojas-Arriagada} et~al.}{2019}]{rojas2019}
{Rojas-Arriagada} A.,  {Zoccali} M.,  {Schultheis} M.,  {Recio-Blanco} A.,
  {Zasowski} G.,  {Minniti} D.,  {J{\"o}nsson} H.,   {Cohen} R.~E.,  2019,
  \mn@doi [\aap] {10.1051/0004-6361/201834126}, \href
  {https://ui.adsabs.harvard.edu/abs/2019A&A...626A..16R} {626, A16}

\bibitem[\protect\citeauthoryear{{Rojas-Arriagada} et~al.,}{{Rojas-Arriagada}
  et~al.}{2020}]{rojas2020}
{Rojas-Arriagada} A.,  et~al., 2020, \mn@doi [\mnras] {10.1093/mnras/staa2807},
  \href {https://ui.adsabs.harvard.edu/abs/2020MNRAS.tmp.2647R} {}

\bibitem[\protect\citeauthoryear{{Schultheis} et~al.,}{{Schultheis}
  et~al.}{2017}]{schultheis2017}
{Schultheis} M.,  et~al., 2017, \mn@doi [\aap] {10.1051/0004-6361/201630154},
  \href {https://ui.adsabs.harvard.edu/abs/2017A&A...600A..14S} {600, A14}

\bibitem[\protect\citeauthoryear{{Spinoso}, {Bonoli}, {Dotti}, {Mayer}, {Madau}
   \& {Bellovary}}{{Spinoso} et~al.}{2017}]{spinoso2017}
{Spinoso} D.,  {Bonoli} S.,  {Dotti} M.,  {Mayer} L.,  {Madau} P.,
  {Bellovary} J.,  2017, \mn@doi [\mnras] {10.1093/mnras/stw2934}, \href
  {https://ui.adsabs.harvard.edu/abs/2017MNRAS.465.3729S} {465, 3729}

\bibitem[\protect\citeauthoryear{{Spitoni}, {Silva Aguirre}, {Matteucci},
  {Calura}  \& {Grisoni}}{{Spitoni} et~al.}{2019}]{spitoni2019}
{Spitoni} E.,  {Silva Aguirre} V.,  {Matteucci} F.,  {Calura} F.,   {Grisoni}
  V.,  2019, \mn@doi [\aap] {10.1051/0004-6361/201834188}, \href
  {https://ui.adsabs.harvard.edu/abs/2019A&A...623A..60S} {623, A60}

\bibitem[\protect\citeauthoryear{{Stanek}, {Udalski}, {Szyma{\'N}ski},
  {Ka{\L}u{\.Z}ny}, {Kubiak}, {Mateo}  \& {Krzemi{\'N}ski}}{{Stanek}
  et~al.}{1997}]{stanek1997}
{Stanek} K.~Z.,  {Udalski} A.,  {Szyma{\'N}ski} M.,  {Ka{\L}u{\.Z}ny} J.,
  {Kubiak} Z.~M.,  {Mateo} M.,   {Krzemi{\'N}ski} W.,  1997, \mn@doi [\apj]
  {10.1086/303702}, \href
  {https://ui.adsabs.harvard.edu/abs/1997ApJ...477..163S} {477, 163}

\bibitem[\protect\citeauthoryear{{Wang} et~al.,}{{Wang}
  et~al.}{2012}]{wang2012}
{Wang} J.,  et~al., 2012, \mn@doi [\mnras] {10.1111/j.1365-2966.2012.21147.x},
  \href {https://ui.adsabs.harvard.edu/abs/2012MNRAS.423.3486W} {423, 3486}

\bibitem[\protect\citeauthoryear{{Wegg} \& {Gerhard}}{{Wegg} \&
  {Gerhard}}{2013}]{wegg2013}
{Wegg} C.,  {Gerhard} O.,  2013, \mn@doi [\mnras] {10.1093/mnras/stt1376},
  \href {https://ui.adsabs.harvard.edu/abs/2013MNRAS.435.1874W} {435, 1874}

\bibitem[\protect\citeauthoryear{{Wegg}, {Gerhard}  \& {Portail}}{{Wegg}
  et~al.}{2015}]{wegg2015}
{Wegg} C.,  {Gerhard} O.,   {Portail} M.,  2015, \mn@doi [\mnras]
  {10.1093/mnras/stv745}, \href
  {https://ui.adsabs.harvard.edu/abs/2015MNRAS.450.4050W} {450, 4050}

\bibitem[\protect\citeauthoryear{{Wegg}, {Rojas-Arriagada}, {Schultheis}  \&
  {Gerhard}}{{Wegg} et~al.}{2019}]{wegg2019}
{Wegg} C.,  {Rojas-Arriagada} A.,  {Schultheis} M.,   {Gerhard} O.,  2019,
  \mn@doi [\aap] {10.1051/0004-6361/201936779}, \href
  {https://ui.adsabs.harvard.edu/abs/2019A&A...632A.121W} {632, A121}

\bibitem[\protect\citeauthoryear{{Wilson} et~al.,}{{Wilson}
  et~al.}{2019}]{wilson2019}
{Wilson} J.~C.,  et~al., 2019, \mn@doi [\pasp] {10.1088/1538-3873/ab0075},
  \href {https://ui.adsabs.harvard.edu/abs/2019PASP..131e5001W} {131, 055001}

\bibitem[\protect\citeauthoryear{{Zana}, {Dotti}, {Capelo}, {Bonoli}, {Haardt},
  {Mayer}  \& {Spinoso}}{{Zana} et~al.}{2018}]{zana2018}
{Zana} T.,  {Dotti} M.,  {Capelo} P.~R.,  {Bonoli} S.,  {Haardt} F.,  {Mayer}
  L.,   {Spinoso} D.,  2018, \mn@doi [\mnras] {10.1093/mnras/stx2503}, \href
  {https://ui.adsabs.harvard.edu/abs/2018MNRAS.473.2608Z} {473, 2608}

\bibitem[\protect\citeauthoryear{{Zasowski} et~al.,}{{Zasowski}
  et~al.}{2013}]{zasowski2013}
{Zasowski} G.,  et~al., 2013, \mn@doi [\aj] {10.1088/0004-6256/146/4/81}, \href
  {https://ui.adsabs.harvard.edu/abs/2013AJ....146...81Z} {146, 81}

\bibitem[\protect\citeauthoryear{{Zasowski}, {Ness}, {Garc{\'\i}a P{\'e}rez},
  {Martinez-Valpuesta}, {Johnson}  \& {Majewski}}{{Zasowski}
  et~al.}{2016}]{zasowski2016}
{Zasowski} G.,  {Ness} M.~K.,  {Garc{\'\i}a P{\'e}rez} A.~E.,
  {Martinez-Valpuesta} I.,  {Johnson} J.~A.,   {Majewski} S.~R.,  2016, \mn@doi
  [\apj] {10.3847/0004-637X/832/2/132}, \href
  {https://ui.adsabs.harvard.edu/abs/2016ApJ...832..132Z} {832, 132}

\bibitem[\protect\citeauthoryear{{Zasowski} et~al.,}{{Zasowski}
  et~al.}{2017}]{zasowski2017}
{Zasowski} G.,  et~al., 2017, \mn@doi [\aj] {10.3847/1538-3881/aa8df9}, \href
  {https://ui.adsabs.harvard.edu/abs/2017AJ....154..198Z} {154, 198}

\bibitem[\protect\citeauthoryear{{Zasowski} et~al.,}{{Zasowski}
  et~al.}{2019}]{zasowski2019}
{Zasowski} G.,  et~al., 2019, \mn@doi [\apj] {10.3847/1538-4357/aaeff4}, \href
  {https://ui.adsabs.harvard.edu/abs/2019ApJ...870..138Z} {870, 138}

\bibitem[\protect\citeauthoryear{{Zoccali} et~al.,}{{Zoccali}
  et~al.}{2003}]{zoccali2003}
{Zoccali} M.,  et~al., 2003, \mn@doi [\aap] {10.1051/0004-6361:20021604}, \href
  {https://ui.adsabs.harvard.edu/abs/2003A&A...399..931Z} {399, 931}

\bibitem[\protect\citeauthoryear{{Zoccali} et~al.,}{{Zoccali}
  et~al.}{2017}]{zoccali2017}
{Zoccali} M.,  et~al., 2017, \mn@doi [\aap] {10.1051/0004-6361/201629805},
  \href {https://ui.adsabs.harvard.edu/abs/2017A&A...599A..12Z} {599, A12}

\makeatother
\end{thebibliography}
\end{document}